\documentstyle[emulateapj]{article}
\begin{document}

\title{FORMATION OF QUASAR NUCLEI IN THE HEART OF ULTRALUMINOUS INFRARED GALAXIES}

\author{Yoshiaki Taniguchi$^1$, Satoru Ikeuchi$^2$, and Yasuhiro Shioya$^1$}

\vspace {1cm}

\affil{$^1$Astronomical Institute, School of Science,
           Tohoku University, Aoba, Sendai 980-8578, Japan}
\affil{$^2$Physics Department, School of Science, Nagoya University, 
           Chikusa, Nagoya 464-8602, Japan}


\begin{abstract}
We investigate whether or not a supermassive black hole (SMBH) 
with mass $\gtrsim 10^8 M_\odot$ can be made in the heart of ultraluminous infrared 
galaxies (ULIGs) during the course of mergers between/among gas rich galaxies.
(a) If one progenitor galaxy had a seed SMBH with mass of $\sim 10^7 M_\odot$,
this seed SMBH can grow up to $\gtrsim 10^8 M_\odot$ due to     
efficient Bondi-type gas accretion during the course of merger given a gas 
density in the circumnuclear region of $n_{\rm H} \sim 10^3$ cm$^{-3}$.
(b) Even if there was no progenitor galaxy with a seed SMBH, star clusters with
compact remnants (neutron stars and/or black holes) produced in the circumnuclear 
starbursts can merge into the merger center within a dynamical
time scale of $\sim 10^9$ years to form a SMBH with $\gtrsim 10^8 M_\odot$.
Note, however, that the contribution of compact remnants supplied from
hidden star clusters is necessary to lead to the formation of a SMBH. 
In conclusion, the ultraluminous infrared galaxies observed in the local universe
can make a SMBH in their center during the course of merger either by
gas accretion onto a seed SMBH or by dynamical relaxation of compact
remnants made in the violent circumnuclear starbursts.
Therefore, it is quite likely that the ULIGs will finally evolve to optically luminous
quasars as suggested by Sanders et al.
\end{abstract} 



\section{INTRODUCTION}

Since the discovery of ultraluminous infrared galaxies 
(Soifer et al. 1984; Wright et al. 1984; see for a review Sanders \&
Mirabel 1996), these galaxies (hereafter ULIGs) have often been 
considered as possible precursors of optically bright quasars
(Sanders et al. 1988a, 1988b; Norman \& Scoville 1988).
This argument is based on the following observational properties of ULIGs.
(a) Their bolometric luminosities amount to $\sim 10^{12} L_\odot$, being 
comparable to those of quasars (Sanders et al. 1988a). (b) Their luminosity
function is similar to that of quasars in the local universe 
(Soifer et al. 1987; Sanders et al. 1988b). 
(c) All the ULIGs are galaxy mergers or heavily 
interacting galaxies (Sanders et al. 1988a; Lawrence et al. 1989; Leech et al. 
1994; Clements et al. 1996). Morphological evidence for galaxy mergers has also
been obtained for a number of optically selected quasars in the local 
universe although the majority are giant elliptical or giant elliptical-like
galaxies (McLeod \& Rieke 1994a, 1994b; Disney et al. 1995; Bahcall et al. 1997;
McLure et al. 1998). However, infrared-selected quasars tend to reside 
in morphologically disturbed hosts (e.g., Hutchings \& Neff 1988; Boyce et al. 1996;
Baker \& Clements 1997).
If most giant elliptical galaxies were formed by 
major mergers between/among disk galaxies (Toomre 1977; Barnes 1989; 
Ebisuzaki, Makino, \& Okumura 1991), it is possible that
the majority of quasar hosts are major merger remnants.  
(d) ULIGs in later merger phases tend to have active galactic nuclei (AGN)
on the average (Sanders et al. 1988b; Majewski et al. 1993; Borne et al. 1997). 
Although all the above properties suggest an evolutionary link from ULIGs 
to optically bright quasars, its plausibility  is still in question.

It is generally considered that the quasars are powered by the central engine
of active galactic nuclei (AGN);
i.e., disk-gas accretion onto a supermassive black hole (SMBH) and
masses of SMBHs in quasar nuclei are estimated
to be $M_\bullet \gtrsim 10^8 M_\odot$ (e.g., Rees 1984; Blandford 1990). 
Therefore, if an evolutionary link exists between 
ULIGs and quasars, we have to explain either the presence or the formation of SMBHs
with mass higher than $\sim 10^8 M_\odot$ in the heart of ULIGs.
This issue was already discussed by Norman \& Scoville (1988). They 
investigated the fate of a coeval, massive-star cluster of $4\times 10^9 M_\odot$
within the central 10 pc region (see also Weedman 1983) and found that
a SMBH can be formed in the heart of ULIGs. However, recent 
high-resolution optical and near-infrared images of a number of ULIGs using 
the Hubble Space Telescope have shown that the intense star forming regions
are scattered in circumnuclear regions up to $\sim$ a few kpc from the nucleus
(Shaya et al. 1994; Scoville et al. 1998; Surace et al. 1998).
Therefore, it still seems  uncertain whether or not a SMBH with
$\gtrsim 10^8 M_\odot$ can be made during the course of merger 
evolution in ULIGs. In this Letter, we investigate this issue taking 
actual observational properties of ULIGs into account. 

\section{FORMATION OF QUASAR NUCLEI IN THE HEART OF ULIGs}

Morphological features of ULIGs suggest that most ULIGs come from
mergers between or among galaxies (Sanders et al. 1988a; Taniguchi, \& Shioya 1998).
Another important property of the ULIGs is that they are very gas rich; 
e.g., $M_{\rm H_2} \sim 10^{10} M_\odot$ (Sanders et al. 1988a; Scoville et al.
1991; Downes \& Solomon 1998). Therefore, the progenitor galaxies of ULIGs are
rich in gas such as giant spiral galaxies.  Since intense starbursts are observed
in many ULIGs, the most probable formation mechanism of SMBHs is the collapse
of compact remnants of massive stars (Weedman 1983; Norman \& Scoville 1988).
Another important issue is whether or not progenitor
galaxies had SMBHs originally in their nuclei.
Although masses of SMBHs in nearly spiral galaxies (i.e., progenitor galaxies
of ULIGs) are of the order of
$10^{6 - 7} M_\odot$ at most (e.g., Kormendy et al. 1998 and references therein),
these seed SMBHs could grow due to gas accretion in circumnuclear dense-gas 
regions during the course of the merger. Therefore, we consider two cases;
1) at least one progenitor had a SMBH with $M_\bullet \sim 10^{6-7} M_\odot$, 
and 2) no progenitor had a SMBH.

\subsection{Mergers between/among Nucleated Galaxies}

If a progenitor galaxy had a SMBH in its nucleus, this seed SMBH could grow
in mass during the course of merger because the central region of a ULIG
is very gas-rich. 
Such gas accretion may be efficient within the central 1 kpc region because
the gas density is quite high in the central region; i.e., $\sim 10^{3-4}$ cm$^{-3}$
(Scoville, Yun, \& Bryant 1997; Downes \& Solomon 1998).
We here consider the classical Bondi-type (Bondi 1952) gas accretion 
onto the SMBH. This gas accretion rate is given by

\begin{equation}
\dot M = 2~\pi ~m_{\rm H} ~ n_{\rm H} ~ r_{\rm a}^2 ~ v_{\rm e},
\end{equation}
where $m_{\rm H}$, $n_{\rm H}$, $r_{\rm a}$, and $v_{\rm e}$ are the mass of
a hydrogen atom, the number density of the hydrogen atom,
the accretion radius defined as $r_{\rm a} = G M_\bullet v_{\rm e}^{-2}$ 
($M_\bullet$ is the mass of the seed SMBH), and the effective relative velocity 
between the seed SMBH and the ambient gas, respectively.
A typical dynamical mass of the central 1 kpc region is of the order of 
$\sim 10^9 M_\odot$.  Suppose that the SMBH with mass of $10^6 M_\odot$ 
is sinking toward the dynamical center of the merger. Its orbital velocity is
$v_{\rm orb} \simeq 67 M_{\rm nuc, 9}^{1/2} r_1^{-1/2}$ km s$^{-1}$,
where $M_{\rm nuc, 9}$ is the dynamical mass of the central 1 kpc region 
of the merger and $r_1$ is the radius in units of 1 kpc. 
The crossing time of the SMBH is $T_{\rm cross} \sim 10^7$ years. 
Therefore, the merging
time scale is estimated to be $T_{\rm merger} \sim 10 T_{\rm cross}
\sim 10^8$ years (e.g., Barnes 1989). 
Adopting an average gas density in the circumnuclear regions of ULIGs
to be $n_{\rm H} \sim 10^3$ cm$^{-3}$,
we obtain the accreting gas mass during the course of merger;

\begin{equation}
M_{\rm acc} = \dot M ~ T_{\rm merger} \sim 3 \times 10^5 M_{\bullet, 6}^2
n_{\rm H, 3} v_{\rm e, 100}^{-3} T_{\rm merger, 8} ~ M_\odot 
\end{equation}
where $M_{\bullet, 6}$ is the mass of SMBH in units of $10^6 M_\odot$, 
$n_{\rm H, 3}$ is the average gas density in units of $10^3$ cm$^{-3}$,
$v_{\rm e, 100}$ is the orbital velocity with respect to the ambient gas
in units of 100 km s$^{-1}$, and $T_{\rm merger, 8}$ is the merger 
time scale in units of $10^8$ years. 
This estimate implies that the seed SMBH cannot grow up
to $M_\bullet \gtrsim 10^8 M_\odot$ if the seed SMBH is less massive than
$10^7 M_\odot$. Since the ULIGs come from 
mergers between two galaxies or among several galaxies, 
their progenitor galaxies should be very giant spiral galaxies
in order to pile up molecular gas up to $\sim 10^{10} M_\odot$ in their 
central regions (e.g., Sanders et al. 1988a).
In fact, nearby spiral galaxies such as M31
and NGC 4258 have SMBHs with a few $10^7 M_\odot$ [e.g., Kormendy et al.
(1998) and references therein; see also Miyoshi et al. (1995)
for the case of NGC 4258]. Therefore, it seems quite likely that
the seed SMBH may be more massive than that adopted in the above estimate.
If the seed SMBH is more massive than a few 
$10^7 M_\odot$, it could grow up to $\gtrsim 10^8 M_\odot$.  
Although we have no knowledge about the seed SMBHs in the progenitors,
our estimates given here suggest that the gas accretion in the dense gas 
clouds onto the seed SMBH can lead to the formation of a quasar nucleus
in the heart of ULIGs.

\subsection{Mergers between/among Non-nucleated Galaxies}

Next we consider a case where there is no seed SMBH in the progenitor
galaxies of ULIGs. In this case, a possible way to form quasar nuclei in ULIGs
is to pile up the circumnuclear star clusters of compact remnants of massive stars;
black holes and neutron stars, each of which has a few $M_\odot$ at most.
This issue was pioneeringly discussed by Weedman (1983).
Using both starburst models by Gehrz, Sramek, \& Weedman (1983) and optical
spectroscopic observations (i.e., H$\alpha$ luminosity), he suggested that 
starburst galaxies with $L$(H$\alpha$) $\gtrsim 10^{42}$ erg s$^{-1}$ could
produce compact starburst remnants up to $\sim 10^9 M_\odot$. 
However, since the H$\alpha$ luminosity of starburst galaxies is dominated by 
the most massive stars in the starburst region, it seems hard to 
estimate the total mass of compact remnants solely using $L$(H$\alpha$).
Thus, the estimate by Weedman (1983) may provide a rough upper limit for the 
compact remnant mass in the starburst galaxies. 
Norman \& Scoville (1988) also discussed the formation of quasar nuclei 
in ULIGs. However, their assumption (a coeval, massive-star cluster of 
$4\times 10^9 M_\odot$ within the central 10 pc region) turns out to be 
unlikely because the recent high-resolution optical and near-infrared imaging
by the Hubble Space Telescope of the ULIGs have shown that blue star clusters
are distributed in the circumnuclear regions up to $r \sim$ a few kpc
(Shaya et al. 1994; Surace et al. 1998; Scoville et al. 1998).
Although the central star cluster associated with the western nucleus of 
Arp 220 is very luminous and its mass is estimated to be $\gtrsim 10^8 M_\odot$,
typical masses of the circumnuclear star clusters are of the order of 
$M_{\rm cl} \sim 10^7 M_\odot$ at most
(Shaya et al. 1994; Scoville et al. 1998; Shioya, Taniguchi, \& Trentham 1998;
see also Taniguchi, Trentham, \& Shioya 1998).
Although some star clusters may be hidden because of heavy extinction
(Scoville et al. 1991; Genzel et al. 1998), 
we firstly investigate whether or not the star clusters 
found both in the optical and in the near infrared (NIR) (Shaya et al. 1994; Scoville 
et al. 1998) will be responsible for the formation of a SMBH with 
$M_\bullet \gtrsim 10^8 M_\odot$.

Shaya et al. (1994) discussed the fate of 
circumnuclear star clusters; since these clusters will lose their kinetic energy
to individual stars during random encounters (i.e., dynamical friction), 
they will sink toward the merger center within $\sim 10^9$ years.
Here, from a viewpoint of dynamical relaxation of the star clusters,
we examine whether or not compact remnants formed in
the circumnuclear star-forming clusters can make a SMBH with 
$M_\bullet \sim 10^8 M_\odot$.
For simplicity, we consider a case where ten circumnuclear star-forming 
clusters, each of which has a total stellar mass of $10^7 M_\odot$, are distributed 
within $r \simeq$ a few kpc. It is known that stars with $m_* \geq 8 M_\odot$ 
produce compact remnants. We estimate how many such massive stars are 
formed in each cluster. 

We assume that stars are formed with a Salpeter-like initial mass function (IMF);

\begin{equation}
\phi(m) = \beta m^{-\mu}
\end{equation}
where $m$ is the stellar mass in units of $M_\odot$ and 
 $\beta$ is a normalization constant determined by the relation

\begin{equation}
\int_{m_l}^{m_u} \phi(m) dm = 1,
\end{equation}
which leads to 

\begin{equation}
\beta  = \frac{(\mu -1)m_l^{\mu-1}}{1 - (m_l/m_u)^{\mu-1}}.
\end{equation}
The number of stars with a mass range $m_1 \le m_* \le m_2$ is estimated as

\begin{equation}
N (m_1 \le m_* \le m_2)= \int_{m_1}^{m_2} \frac{\phi(m)}{m} dm.
\end{equation}
Using $\beta$ in equation (5), we re-write equation (6) as 

\begin{equation}
N (m_1 \le m_* \le m_2) = \left( \frac{\beta}{\mu} \right)
(m_1^{- \mu} - m_2^{- \mu}) ~~ {\rm stars}~M_\odot^{-1}.
\end{equation}
There are three free parameters; the power index ($\mu$), and the upper and
lower mass limits of the IMF ($m_u$ and $m_l$). Since stars with $m_* \geq 8 M_\odot$
produce compact remnants, we set $m_1 = 8 M_\odot$ and $m_2 = m_u > m_1$.
In Table 1, we give results for some possible combinations of the parameters. 
Although $\mu$ = 1.35 is the canonical value for stars in the solar neighborhood,
there are some lines of evidence that massive stars
are more preferentially formed in such violent star-forming regions;
i.e., a top-heavy initial mass function with $m_l \sim 1 M_\odot$
(e.g., Goldader et al. 1997 and references therein).
Therefore, we adopt a case of $\mu = 0.35$, $m_u = 30 M_\odot$ and 
$m_l = 1 M_\odot$. In this case, there are $\sim 4 \times 10^5$ massive stars 
with $m_* \geq 8 M_\odot$ in each cluster. Each compact remnant has a mass from 
a few $M_\odot$ (for neutron stars) to several  $M_\odot$ (for black holes). 
Therefore, the total mass of compact remnants in each cluster is 
$\sim 10^6 M_\odot$.
Such ten clusters will be relaxed dynamically with a time scale of

\begin{equation}
T_{\rm dyn} \sim n_{\rm cl}^{1/2} G^{-1/2} r_{\rm cl}^{3/2} M_{\rm cl}^{-1/2}
\sim 9.4 \times 10^8 r_{\rm cl, 1}^{3/2} M_{\rm cl, 7}^{-1/2} ~~{\rm (years)},
\end{equation}
where $r_{cl, 1}$ is the typical size of a circumnuclear region with 
ten star clusters in units of 1 kpc and 
$M_{\rm cl, 7}$ is the mass of the clusters in units of $10^7 M_\odot$.
Therefore, we expect that a SMBH with mass of $\sim 10^7 M_\odot$ will be made
$\sim 10^9$ years after the onset of circumnuclear starbursts in ULIGs.
This mass is smaller than that necessary for quasar nuclei 
(i.e., $\gtrsim 10^8 M_\odot$). Each cluster of compact remnants would be able 
to gain its mass by gas accretion as discussed in section 2.1. However,
the accreted mass is estimated to be $\sim 3 \times 10^7 M_\odot$ for 
ten clusters in total, being still less massive  than $10^8 M_\odot$.

Here it should be again remembered that all star clusters in the central region of
ULIGs cannot be observed in both the optical and the NIR because 
inferred extinction toward the nuclei of ULIGs is very large; e.g.,
$\gtrsim$ 50 mag (Genzel et al. 1998; Scoville et al. 1991). Therefore,
it is quite likely that the majority of nuclear star clusters in ULIGs 
are hidden by a large amount of gas and dust. 
Recently, Shioya et al. (1998) analyzed the optical-NIR spectral energy
distributions of nuclear star clusters in Arp 220. They found that 
these clusters are more massive systematically (i.e., $\gtrsim 10^8 M_\odot$)
than circumnuclear ones but can account only for about one-seventh of
the total bolometric luminosity of Arp 220.  Although OH megamaser sources found
in the central region of Arp 220 may provide possible evidence for  hidden AGN
(Diamond et al. 1989; Lonsdale et al. 1998), the recent mid-infrared spectroscopy
of a sample of ULIGs has shown that the majority of the ULIGs such as Arp 220
are powered by nuclear starbursts (Genzel et al. 1998; Lutz et al. 1998).
Therefore, it is strongly suggested that some hidden star clusters should be 
responsible for the remaining ($\sim 6/7$) bolometric luminosity of Arp 220. 
Since compact remnants produced in the hidden clusters will join to form a SMBH,
it is expected that a SMBH with $M_\bullet \gtrsim 10^8 M_\odot$ will be formed
in the heart of ULIGs.

\section{DISCUSSION}

We have shown that a SMBH with mass $\gtrsim
10^8 M_\odot$ can be made in a 
ULIG during the course of merger between/among gas rich galaxies
regardless of the presence of a seed SMBH in progenitor galaxies; i.e., 
(a) if one progenitor galaxy had a seed SMBH with mass of $\sim 10^7 M_\odot$,
this seed SMBH can grow up to $\gtrsim 10^8 M_\odot$ because of
efficient Bondi-type gas accretion during the course of merger given 
a gas density in the circumnuclear region of $n_{\rm H} \sim 10^3$ cm$^{-3}$, and 
(b) even if there was no progenitor galaxy with a seed SMBH, the star clusters of 
compact remnants made in the circumnuclear starbursts can merge into
the merger center within a dynamical time scale of $\sim 10^9$ years
to form a SMBH with $\gtrsim 10^8 M_\odot$.
Note, however, that the contribution of compact remnants supplied from
hidden star clusters is necessary to explain the formation of SMBHs.
In conclusion, the ultraluminous infrared galaxies observed in the local universe
can make a SMBH in their center during the course of merger either by
the gas accretion onto a seed SMBH or by the dynamical relaxation of
star clusters of  compact remnants made in the violent circumnuclear starbursts.

The presence of a SMBH  with $M_\bullet \gtrsim 10^8 M_\odot$
is a crucially important necessary condition 
for the occurrence of quasar activity. 
In the local universe, the masses of SMBHs in giant spiral galaxies
(e.g., our Milky Way Galaxy, M31, NGC 1068, and NGC 4258) are as low as
$M_\bullet \sim 10^{6-7} M_\odot$  (Kormendy et al. 1998 and references therein).
Therefore, it is suggested that isolated, typical spiral galaxies
cannot harbor quasar nuclei. However, 
mergers between/among gas-rich galaxies can cause efficient gas fueling
toward the nuclear regions of the merging systems and then trigger
intense starbursts either as a result of the piling of a lot of gas
(Mihos \& Hernquist 1994) or by the dynamical effect of SMBH
binaries (Taniguchi \& Wada 1996; Taniguchi \& Shioya 1998).
Furthermore, as demonstrated in the present work, these mergers
provide a possible way to form SMBHs with $M_\bullet \gtrsim 10^8 M_\odot$.
In this respect, 
it is quite likely that the ULIGs will finally evolve to optically luminous
quasars  as suggested by Sanders et al. (1988a, 1988b).

Finally it is worthwhile noting that some elliptical galaxies 
could be formed by galaxy mergers (e.g., Toomre 1977; Barnes 1989;  
Ebisuzaki et al. 1991).
We also note that some elliptical galaxies
(e.g., M87, NGC 3115, NGC 3377, NGC 4261, and so on)  have
SMBHs with $M_\bullet \gtrsim 10^8 M_\odot$ (Kormendy et al. 1998
and references therein). Actually, investigating the physical
conditions of ULIGs (i.e., mass density and velocity dispersion), 
Kormendy and Sanders (1992) found evidence that ULIGs are elliptical galaxies
forming by merger-induced dissipative collapse (see also Wright et al. 1990;
Baker \& Clements 1997). 
Therefore, we suggest that almost all SMBHs with $M_\bullet \gtrsim 10^8 M_\odot$
in the local universe were made by galaxy mergers.

\vspace{0.5cm}

We would like to thank an anonymous referee for his/her useful comments.
This work was supported in part by the Ministry of Education, Science,
Sports and Culture in Japan under Grant Nos. 07055044, 10044052, and 10304013.

\newpage


\newpage

\begin{table}
\caption{The number of massive stars per unit mass ($M_\odot$)}
\begin{tabular}{ccccc}
\tableline
\tableline
$\mu$ & $m_l$ & $m_u$ & $\beta$ & $N(m_* \geq 8 M_\odot)$ \\
      & $(M_{\odot})$ & $(M_{\odot})$ & & (stars $M_\odot^{-1}$) \\
\tableline
0.35 & 1 & 100 & 0.034 & 0.0278 \\
0.35 & 1 &  30 & 0.080 & 0.0409 \\
1.35 & 1 & 100 & 0.437 & 0.0189 \\
1.35 & 1 &  30 & 0.503 & 0.0187 \\
\tableline
\end{tabular}
\end{table}

\end{document}